\documentclass{elsart4-1}

\usepackage{epsfig}

\usepackage{amssymb}
\usepackage{amsmath}

\usepackage[english,francais]{babel}

\newtheorem{e-proposition}[theorem]{Proposition}

\newtheorem{e-definition}[theorem]{Definition\rm}

\def\og{\leavevmode\raise.3ex\hbox{$\scriptscriptstyle\langle\!\langle$~}}
\def\fg{\leavevmode\raise.3ex\hbox{~$\!\scriptscriptstyle\,\rangle\!\rangle$}}

\begin{document}

\begin{frontmatter}


\selectlanguage{english}
\title{The Ruthenocuprates - Natural Superconductor-Ferromagnet Multilayers}


\selectlanguage{english}
\author[1]{Timo Nachtrab}
\author[2]{Christian Bernhard}
\author[2]{Chengtian Lin}
\author[1]{Dieter Koelle}
\author[1]{Reinhold Kleiner}
\ead{kleiner@uni-tuebingen.de}

\address[1]{Physikalisches Institut-Experimentalphysik II, Universit\"at T\"ubingen, Auf der Morgenstelle 14, 72076 T\"ubingen, Germany}
\address[2]{Max-Planck-Institut f\"ur Festk\"orperforschung, Heisenbergstr. 1, 70569 Stuttgart, Germany}


\medskip
\begin{center}
\end{center}

\begin{abstract}
The recently discovered ruthenocuprates have attracted great interest because of the microscopic coexistence of superconducting and ferromagnetic order. Typically, these materials become magnetically ordered at temperatures around 125-145\,K and superconductivity sets in between 15 and 50\,K. While superconductivity arises in the CuO$_2$ layers the RuO$_2$ layers in between order magnetically. In this paper we summarize some of the crystallographic, magnetic and superconducting properties of the ruthenocuprates, as obtained from investigations on polycrystalline samples as well as single crystals. 

\vskip 0.5\baselineskip

\selectlanguage{francais}
\noindent{\bf R\'esum\'e}
\vskip 0.5\baselineskip
\noindent
D\'ecouverts r\'ecemment, les ruthenocuprates ont suscit\'e un int\'er\^et \'enorme \`a cause de la coexistence d'un ordre supraconducteur et ferromagn\'etique. Dans ces mat\'eriaux, la phase magn\'etique appara\^it autour de 125-145\,K et la phase supraconductrice entre 15 et 50\,K. Les couches de CuO$_2$ sont supraconductrices et ce sont les couches interstitielles de RuO$_2$ qui sont ordonn\'ees magn\'etiquement. Dans cet article nous r\'esumons quelques-unes des propri\'et\'es cristallographiques, magn\'etiques et supraductrices des ruthenocuprates d\'eduites des \'etudes sur des \'echantillons polycristallins ainsi que sur des monocristaux. 

 \vskip 0.5\baselineskip
\noindent{\small{\it Mots-cl\'es~:} supraconductivit\'e~; magn\'etisme~;
ruthenocuprate~; multicouches}}
\end{abstract}
\end{frontmatter}


\selectlanguage{english}
\section{Introduction}
\label{}
During the last years, and particularly after the discovery of high-$T_c$ superconductivity in cuprates in 1986, the physics of doped oxides has become a subject of enormous research activities \cite{Ima98}. These systems with their complex phase diagrams and their subtleties concerning dopants and doping levels allow to investigate the interplay between orbital physics, structural properties like lattice distortions and the correlation of charge carriers (e.g. charge ordering). A special class of doped oxides are the various types of naturally layered, conducting compounds with their highly anisotropic, quasi two-dimensional crystal structure. A prominent example are the high-$T_c$ cuprates where superconducting CuO$_2$ sheets alternate with block layers that may be normal conducting or even insulating. For example, in the compound Bi$_2$Sr$_2$CaCu$_2$O$_8$ (BSCCO) CuO$_2$ double layers are separated by SrO and BiO planes. Interlayer charge transport occurs via tunneling between adjacent CuO$_2$ double layers. In the superconducting state interlayer supercurrents flow as Josephson currents making the compound a natural stack of Josephson junctions (intrinsic Josephson effect) \cite{Kle92,Yur00}. 
\begin{figure}[h]
\centering
\epsfig{file=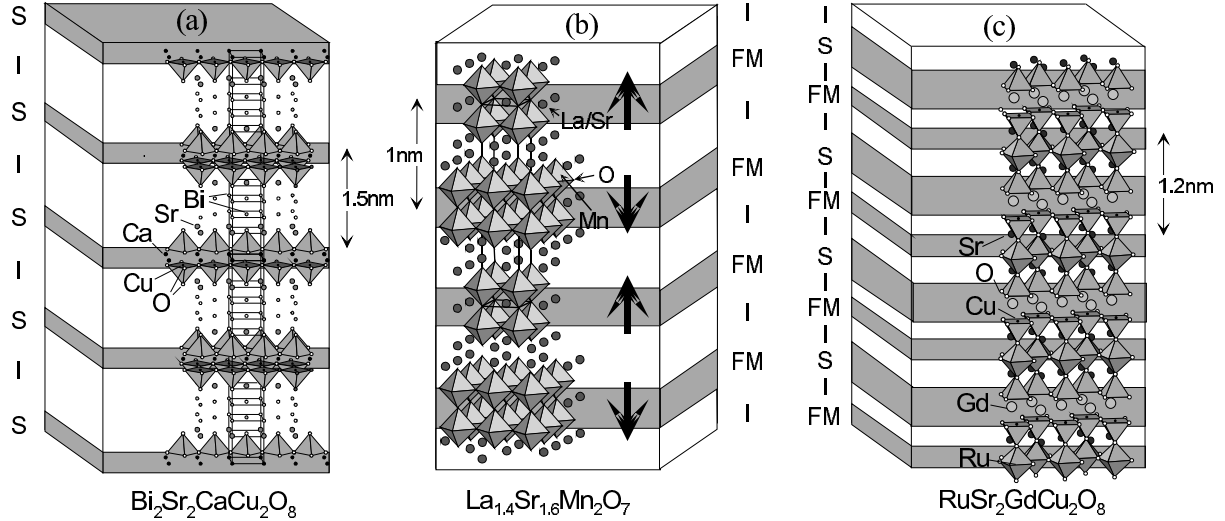}
\caption{Crystal structures of Bi$_2$Sr$_2$CaCu$_2$O$_8$ (a), La$_{1.4}$Sr$_{1.6}$Mn$_2$O$_7$ (b) and RuSr$_2$GdCu$_2$O$_8$ (c) superposed to the model of, respectively, a superconductor-insulator-superconductor (SIS) multilayer, a ferromagnet-insolator-ferromagnet (FM-I-FM) multilayer and a FM-I-S-I-FM multilayer. Arrows in (b) denote orientation of magnetization.}
\end{figure}
Figure 1(a) shows the crystallographic structure of this compound superposed to a schematic of a SIS superlattice (S and I, respectively, denote superconducting and insulating layers). Another example for a layered conducting material is the layered manganite \cite{Kim96} La$_{1.4}$Sr$_{1.6}$Mn$_2$O$_7$ (LSMO), cf. figure 1(b). It consists of MnO$_3$ bilayers separated by layers of (La,Sr)O. The interplay between the charge carriers in LSMO is of the double-exchange type, leading to an insulator-to-metal transition inside the MnO$_3$ bilayers near 100\,K, accompanied by a transition from a paramagnetic high temperature phase to a ferromagnetic low temperature phase \cite{Kim96}. Like for BSCCO, the layers of SrO in between act as a tunneling barrier for the (in this case spin-polarized) current along the $c$-axis. Below about 75\,K  adjacent (ferromagnetic) MnO$_3$ bilayers order antiferromagnetically with respect to each other, the easy axis oriented along the $c$-axis. The interlayer tunnelling conductivity depends on the relative orientation of the magnetizations at both sides of the barrier, being low for antiparallel orientation and large for parallel orientation. An external magnetic field can be used to rotate magnetization vectors towards parallel alignment, resulting in a large negative magnetoresistance. The material thus exhibits an "intrinsic spin valve effect" \cite{Wel99,Nac01}.

The ruthenocuprates combine properties of the above systems. As we will see below there are compounds of the type RuSr$_2$($R_{1+x}$Ce$_{1-x}$)Cu$_2$O$_{10}$, with $R$ = Sm, Eu, Gd, and compounds of the type RuSr$_2R$Cu$_2$O$_8$ , with $R$ = Gd, Eu, Y. The former class is usually referred to as Ru-1222 while the latter class is referred to as Ru-1212. In order to specify $R$, sometimes this element is placed after this notation, e. g. Ru-1212Gd for RuSr$_2$GdCu$_2$O$_8$. We will use this notation within this article.
 
Ru-1212Gd consists of an alternating sequence of weakly ferromagnetic (F), insulating (I), and superconducting (S) sheets along the $c$-axis, cf. figure 1(c). As we will see, in the superconducting state the material exhibits an intrinsic Josephson effect while in the normal state a negative magnetoresistance can be observed. For such an intrinsic SIFIS system  a variety of new effects may be expected. For example, the magnetization of the weakly ferromagnetic sheets should provoke screening currents in the superconducting subsystem, generating a spontaneous vortex phase \cite{Fel97,Son98,Pic99,Ber00,Son02}. Due to the exchange splitting in the CuO$_2$ layers arising from the magnetic subsystem the Cooper pairing may occur with finite momenta, leading to a spatially modulated superconducting order parameter \cite{Pic99,Zhu00,Shi00}. This so-called Fulde-Ferrell-Larkin-Ovchinnikov (FFLO) state \cite{Ful60,Lar65} may have a wave vector $\vec{q}$ parallel to the layers, roughly pointing in (110) direction \cite{Shi00}. To obtain this state the exchange splitting must be sufficiently large (of order $\Delta/2$, where $\Delta$ is the superconductor energy gap \cite{Zhu00}). If this condition is not fulfilled still the exchange field inside the ferromagnetic sheets may cause the superconducting order parameter to change sign between adjacent superconducting layers. As a consequence, having Josephson coupling between CuO$_2$ double layers, there would be an additional phase shift of $\pi$ in the Josephson current-phase relation, $I=I_c \sin(\varphi+\pi)$, resulting in intrinsic Josephson $\pi$-junctions \cite{Pic99,Hou01,Pro99}. Here, $\varphi$ denotes the difference of the phases of the superconducting order parameters of the two superconducting layers and $I_c$ is the maximum supercurrent across the layers. Further the possibility of triplet Cooper pairing was proposed \cite{Tal00}. Note that the ruthenate SrRuO$_4$ is such a triplet superconductor \cite{Mac03}.

In this paper some of the properties of the ruthenocuprates will be discussed. There are several hundred papers on these materials and the number is growing rapidly. Although numerous investigations have been performed on both Ru-1212 and on Ru-1222 many issues regarding both magnetic and superconducting ordering are still under debate. We thus by no means can take full account of all developments and will focus on some selected issues.

In the next section some of the basic properties of Ru-1212 and Ru-1222 (crystal structure, superconducting and magnetic ordering) will be introduced. Most experiments so far have been made with polycrystalline samples and thus this section will focus on results obtained from such samples, with a certain focus on Ru-1212. In section \ref{singlecrystals} we will turn to Ru-1212Gd single crystals and discuss in some detail results of magnetization and interlayer transport experiments performed with these single crystals. Conclusions are given in section \ref{conclusions}.

\section{ The Ruthenocuprates - Some Basic Properties}
\label{basicproperties}
In 1995 the first ruthenocuprates of, respectively, the Ru-1212 and the Ru-1222 type were synthesized by Bauernfeind et al. \cite{Bau95,Bau96}. The idea was to insert metallic layers into the layered structure of a high-$T_c$ cuprate in order to increase the critical current density of these compounds. 
\begin{figure}[h]
\centering
\epsfig{file=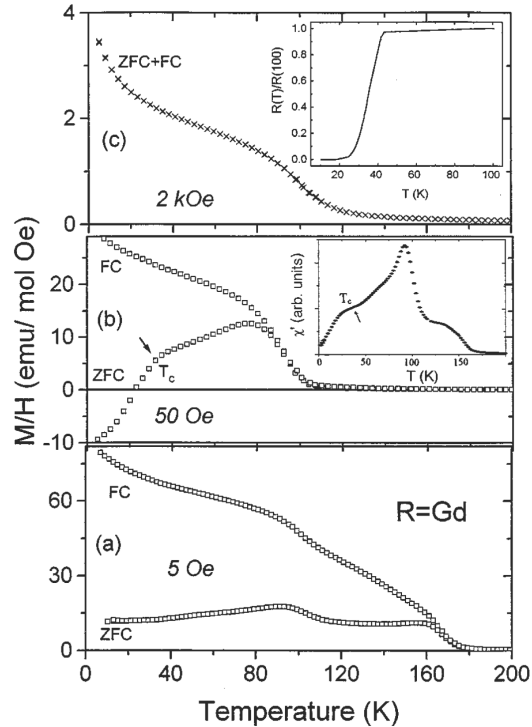}
\caption{Magnetization data of Gd$_{1.4}$Ce$_{0.6}$RuSr$_2$Cu$_2$O$_{10}$ showing the coexistence of superconductivity and magnetism. Measurements were taken in fields of (a) 5\,Oe, (b) 50\,Oe and (c) 2\,kOe. Inset in (c) shows resistance vs. temperature, inset in (b) shows the ac susceptibility vs. temperature. FC denotes field cooled measurements, ZFC zero field cooled measurements (from \cite{Fel97}). }
\end{figure}
The RuO$_2$ layers have the same square-planar coordination and a similar bond length as CuO$_2$ and are thus suitable to be inserted into the cuprates. Subsequent studies revealed that this class of materials exhibit ferromagnetic order at Curie temperatures well above the superconducting transition temperature \cite{Fel97,Fel00,Ber99,Tal99}. The magnetism persists in the superconducting state. As a first example, figure 2 shows magnetization data of Felner {\it et al.} \cite{Fel97} for Gd$_{1.4}$Ce$_{0.6}$RuSr$_2$Cu$_2$O$_{10}$. The plots correspond to measurements in three magnetic fields of 5\,Oe (a), 50\,Oe (b) and 2\,kOe (c). In (a) there are magnetic field anomalies at about 102\,K and 170\,K attributed to the Ru ions. Below about 42\,K the sample becomes superconducting. Felner {\it et al.} found via magnetic susceptibility and M\"ossbauer spectroscopy that superconductivity seems to be confined to the CuO$_2$ planes whereas the magnetism is due to the Ru sublattice.

As another example we show early data \cite{Ber99,Tal99} for Ru-1212Gd. Figure 3 shows magnetization data. 
\begin{figure}[h]
\centering
\epsfig{width=6cm,file=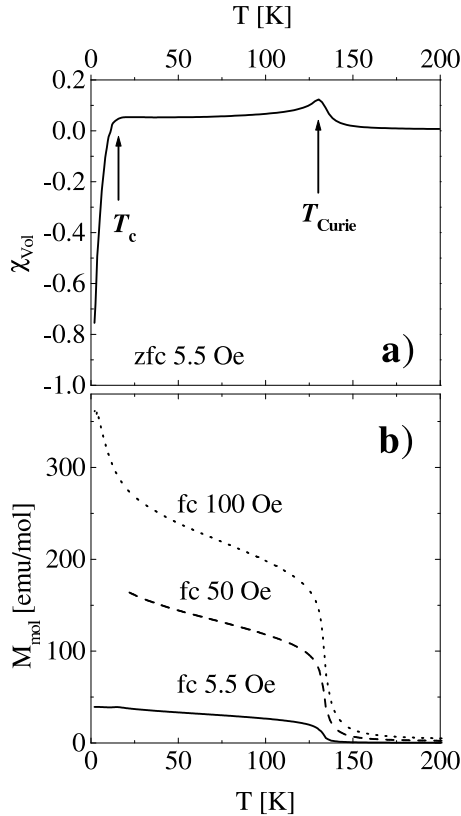}
\caption{Temperature dependence of the zero field cooled volume magnetization $\chi_{Vol}$ (a) and the field cooled molar magnetization $M_{mol}$ (b) for Ru-1212Gd. Arrows in (a) denote the ferromagnetic transition at 133\,K and the superconducting transition at 16\,K (from \cite{Ber99}).}
\end{figure}
This particular sample became superconducting at 16\,K while magnetic order already set in at 133\,K. More generally, for Ru-1212, depending on preparation conditions and annealing, $T_c$ can vary in the range 15-50\,K, the strong dependence perhaps caused by the defect structure of the materials \cite{Vas04}. Muon spin resonance revealed that the magnetic order is a bulk effect and that it persists on a microscopic scale also in the superconducting state. The corresponding data are shown in figure 4 for the same Ru-1212Gd sample with $T_c = 16$\,K. 
\begin{figure}[h]
\centering
\epsfig{file=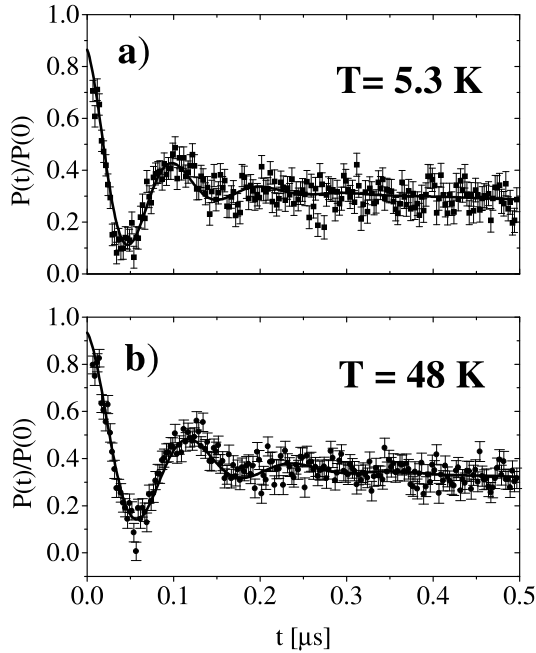}
\caption{Normalized time resolved muon-spin polarization $P(t)/P(0)$ of a Ru-1212Gd sample with $T_c = 16$\,K for temperatures of (a) 5.3\,K (i. e. in the superconducting state) and (b) 48\,K (i. e. in the normal conducting state). The large oscillatory component of $P(t)$ gives evidence for the presence of a bulk magnetically ordered state (from \cite{Ber99}). }
\end{figure}
In the graph the normalized time resolved muon-spin polarization $P(t)/P(0)$ is plotted for temperatures of 48\,K and 5.3\,K. The large oscillatory component of $P(t)$ gives evidence for the presence of a bulk magnetically ordered state and is observable both in the normal state and in the superconducting state. For the same samples measurements of the specific heat exhibit a pronounced jump at $T_c$ and thus establish that superconductivity is indeed also a bulk effect \cite{Tal00}.
As we will see below the question whether or not superconductivity truly coexists with magnetism, or if some phase separation occurs on a nanoscale, is under debate. We thus already note here that, recently, muon spin resonance experiments similar to the one discussed above have also been performed on Eu$_{1.4}$Ce$_{0.6}$RuSr$_2$Cu$_2$O$_{10}$ showing that also in this compound magnetism is present on a microscopic scale in the superconducting state \cite{She04}. For Ru-1212 the true coexistence of superconductivity and magnetism has also been concluded from SQUID magnetometer measurements using stationary samples, thus avoiding parasitic effects arising in magnetization measurements where the sample is moved \cite{Pap03,Pap02}.

\subsection{Synthesis and Crystal Structure}
Polycrystalline ruthenocuprate samples can be obtained by a solid-state reaction. For example, to obtain Ru-1212Gd one starts from stoichiometric powders of RuO$_2$, SrCO$_3$, Gd$_2$O$_3$ and CuO. The mixture is first decomposed at around 960$^\circ$C, then ground, die-pressed and sintered at 1010$^\circ$C. Afterwards further sintering (at 1050$^\circ$C) and annealing steps are required to obtain single phase polycrystals, as revealed by x-ray diffraction \cite{Ber99}.

Single crystals of Ru-1212Gd have been grown by the self-flux method \cite{Lin01}. Unfortunately, Ru-1212Gd melts incongruently and crystals form at relatively high temperatures above 1100$^\circ$C where it is difficult to achieve a significant degree of solubility of Ru atoms in the crystal because of the very high Ru vapor pressure and escape rate. Nonetheless good crystals were obtained using a solvent Ru:Gd:Cu = 0.8:0.4:1.7 and a low cooling rate of 1.5$^\circ$C/h. The crystals obtained - figure 5 shows SEM images - are very small, with typical sizes $200 \times 200 \times 50$\,$\mu$m$^3$ or even smaller, making a variety of experiments difficult or even impossible.
\begin{figure}[h]
\centering
\epsfig{width=9cm,file=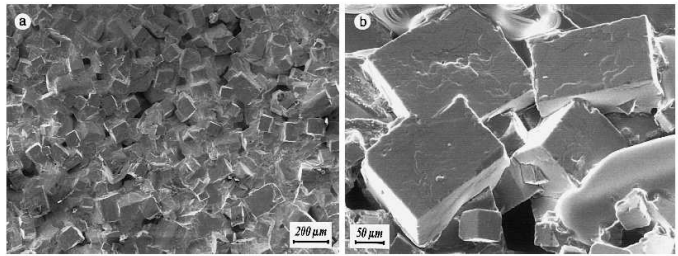}
\caption{SEM images of typical single crystals of Ru-1212Gd (from \cite{Lin01}).}
\end{figure}
Recently, also Ru-1222 single crystals have been grown \cite{Wat04}, as well as melt textured Ru-1212Gd samples \cite{Gom04}. Also, recently Ru-1212Gd thin films have been made using pulsed laser deposition \cite{Leb05,Mat04}.

The Ru-1212Gd crystal structure is comparable to that of YBa$_2$Cu$_3$O$_7$ (YBCO), with Y and Ba replaced by Gd and Sr, respectively. The Cu ions of the CuO chains are replaced by Ru ions \cite{Chm00,McL99}. The compound then consists of CuO$_2$ double layers, the cooper ions being surrounded by tetrahedrons of oxygen ions (cf. figure 1c). The Ru ions are located in the middle of oxygen octahedrons alternatingly rotated by 14$^\circ$ clockwise or counterclockwise, respectively (cf. figure 6). 
\begin{figure}[h]
\centering
\epsfig{file=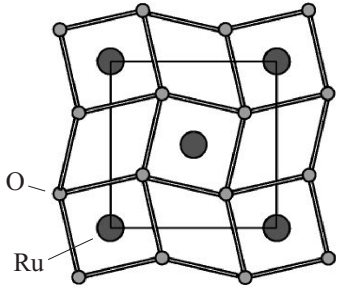}
\caption{Top view of a RuO$_2$ layer. The octahedrons are rotated along the $c$-axis (from \cite{Nak01}).}
\end{figure}
As we will see below this rotation is important to understand the magnetic ordering of the Ru ions. The space group of Ru-1212Gd is P4/mbm, with dimensions $a/b \approx 5.43$\,\AA\ and $c \approx 11.56$\,\AA. Neglecting the rotation of the octahedra the more simple space group is P4/mmm. Here the dimensions of the unit cell are $a/b \approx 3.83$\,\AA\ and $c \approx 11.56$\,\AA.

In comparison to Ru-1212Gd in the Ru-1222 compound the Gd layer between the CuO$_2$ sheets is replaced by a double flourite-type ($R_{1-x}$Ce$_x$)$_2$O$_2$ block \cite{Fel97,McL03}. The Ru sublattice remains unchanged. 

\subsection{Ru-1212: Electronic properties of the Ru sublattice}
In early reports on Ru-1212 it was suggested that the ruthenium ions exist as Ru$^{5+}$ with a low-spin state configuration ($S = 1/2$) \cite{Ber99}. However, further results from x-ray absorption \cite{Liu01} and NMR studies \cite{Tok01,Kum01} indicated that the Ru ions appear as a mixture of 40\% Ru$^{4+}$ (low-spin, $S = 1$) and 60\% Ru$^{5+}$ (high-spin, $S = 3/2$), being antiferromagnetically coupled via the superexchange mechanism. The mixed valence is most likely caused by a charge transfer from the CuO$_2$ sheets to the RuO$_2$ layers \cite{Kum01,Man02}. Due to this mixed valence a double exchange mechanism inside the RuO$_2$ sheets was proposed \cite{Nak02,Ali04}, mediated by the small canting of the Ru octahedra \cite{But01}. Indeed, from Hall effect and thermopower measurements it was concluded that the conductivity inside the RuO$_2$ layers increases strongly below the temperature of the magnetic ordering \cite{McC03}. While at room temperature the conductivity of these layers is at most 10\% of that of the CuO$_2$ layers, below T$_{mag}$ it rises to at least 30\% of the conductance $\sigma_{Cu}$ of the CuO$_2$ layers (the latter appearing to be similar to the conductance of the high $T_c$ cuprates). Metallic behavior of the RuO$_2$ layers was also inferred in \cite{Liu05,Poz02} based on measurements of magnetothermopower, magnetoresistivity and microwave absorption and dispersion. Doping of the CuO$_2$ layers thus occurs by band overlap which is rather rare for cuprate materials \cite{McL03}. It should, however be noted here that $^{99}$Ru M\"ossbauer effect measurements on Ru-1212Gd revealed only a single Ru site with a valency of +5 \cite{Mar02}. Further, a NMR and x-ray diffraction study on Ru-1212Eu revealed that, while the Ru moments indeed have a ferromagnetic component the Ru 4$d$ electrons are localized \cite{Sak03}. The conclusion is based on the fact that for itinerant ferromagnets the temperature dependence of the unit cell volume should be very weak (Invar effect). By contrast the unit cell volume of Ru-1212Eu changes significantly between 300\,K and 4\,K.

Regarding Ru-1222, x-ray measurements on Ru-1222Gd showed that the average Ru valence is between 4.95 and 5 irrespective of the Ce concentration \cite{Fel99,Wil02}. Further, also the charge transfer to the CuO$_2$ layers due to Ce-doping seems to be low. Nonetheless, Raman and magnetization studies \cite{Wil02}, as well as ac susceptibility, Ru 3$p$ x-ray photoemission and Ru 2$p$ x-ray absorption measurements of Ru-1222Eu suggested that the RuO$_2$ layers are metallic at all temperatures \cite{Hir02}. 

\subsection{Ru-1212: Magnetic structure of the Ru sublattice}
One of the points that are still discussed controversially is the magnetic ordering of the Ru ions. Some of the results obtained by magnetization measurements, neutron diffraction, NMR or ESR studies seem to be in contradiction with each other. In this subsection we focus mainly on Ru-1212Gd. Results for Ru-1222 will be given further below.

For Ru-1212Gd the Ru sublattice orders at about 133\,K. Measurements like muon spin resonance ensure that the magnetic ordering is indeed a property of the whole system and not only of a parasitic phase. Based on magnetization and $\mu$SR measurements the early publications gave evidence for a ferromagnetic ordering of the Ru sublattice \cite{Ber99,Tal99} with the easy axis perpendicular to the $c$-axis (i.e. along the planes). The estimated magnetic moment per Ru ion was $\mu_{Ru} \approx 1.05\mu_B$. Neutron diffraction measurements of various groups, however, showed that the Ru sublattice orders predominantly antiferromagnetically along all crystallographic directions (G-type antiferromagnetism) with the easy axis oriented perpendicular to the layers. For the magnetic moment values $\mu_{Ru} \approx 1.18\mu_B$ \cite{Chm00,Lyn00,Jor01,Tak01} were found. If there was a ferromagnetic component in-plane, it would have a net (Ru plus Gd ions) upper limit of $0.1\mu_B$ \cite{Lyn00}. In this scenario, for external magnetic fields applied perpendicular to the planes a spin-flop like transition, accompanied by a sudden increase of the ferromagnetic signal, is to be expected. Indeed neutron diffraction measurements by Lynn {\it et al.} \cite{Lyn00} showed such an increase above an external field of roughly 0.4\,T. On the theoretical side, while first calculations by Weht {\it et al.} \cite{Weh99,Pic99} that neglected the above mentioned rotation of the RuO$_6$ octahedrons resulted in a ferromagnetic alignment of the Ru ions, calculations by Nakamura {\it et al.} showed that the Ru lattice orders indeed antiferromagnetically (G-type) when taking into account this kind of distortions \cite{Nak01}.

The question that remains is how the antiferromagnetism can be brought into accordance with the observation of ferromagnetic signals in magnetization measurements. A possible explanation, as it has been given by Jorgensen et al. \cite{Jor01}, is shown in figure 7.
\begin{figure}[h]
\centering
\epsfig{file=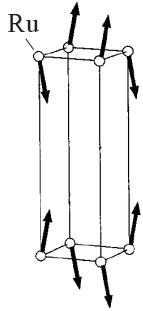}
\caption{Proposed magnetic structure of the Ru sublattice in Ru-1212 that would explain the results from neutron diffraction experiments. The dominant ordering is G-type antiferromagnetic. In addition the local magnetic moments are slightly canted along the $ab$-planes, resulting in a small ferromagnetic component (from \cite{Jor01}).}
\end{figure}
While the dominant ordering of the Ru lattice is antiferromagnetic, with the easy axis oriented perpendicular to the layers, the whole subsystem is slightly canted resulting in a net ferromagnetic component along the $ab$-planes. Comparing Ru-1212Gd and Ru-1212Eu where the magnetic Gd ions have been replaced by nonmagnetic Eu the authors estimate the ferromagnetic component of the Ru sublattice to be about $0.034\mu_B$. Including the Gd ions the net ferromagnetic component is estimated to be about $0.2\mu_B$. Calculations by Nakamura {\it et al.} \cite{Nak02} taking into account a small canting of the Ru ions support the picture of a canted antiferromagnetic order. In their scenario the magnetic moment of the Ru ions projected to the (antiferromagnetic) $c$-axis is about $1.16\mu_B$. The projected moment along the ferromagnetic axis is about $0.99\mu_B$, i. e. quite larger than the estimate of $0.034\mu_B$ given above.
 
It should also be stressed that there are results from ESR \cite{Fai99} and NMR \cite{Tok01} experiments indicating that the coupling between the Ru ions is of the ferromagnetic type, with the magnetic moments lying in-plane. Taking into account the quasi 2-D structure of the RuO$_2$ planes, as well as a small interplane coupling, an alternative model was proposed \cite{But01} where the magnetic ordering of the Ru ions is ferromagnetic inside the RuO$_2$ sheets with the easy axis lying in-plane, while the inter-plane coupling is antiferromagnetic (Type I antiferromagnet), which would make the magnetic order similar to the layered manganites mentioned in the introduction. 

\subsection{Ru-1212Gd: Magnetic Structure of Gd Sublattice} 
The Gd ions have a magnetic moment of about $7\mu_B$ \cite{Ber99}.  Magnetic moments that large strongly influence many of the above mentioned measurements, and thus a lot of experiments have been performed on systems where Gd was replaced by Eu or Y. Thus, the Gd lattice is not investigated in such detail as the Ru lattice. From neutron diffraction measurements, it is known, however, that the Gd ions order antiferromagnetically (G-type) at about 2.6\,K, with their magnetic moments along the $c$-axis \cite{Lyn00}.

\subsection{Ru-1222: Magnetic Structure} 
Although the crystal structure of Ru-1222 compounds is similar to that of Ru-1212 its magnetic and electronic structure seems to exhibit several differences. We already mentioned the two magnetic phase transitions at near 170\,K and 100\,K. While the latter is likely to correspond to the magnetic ordering of the RuO$_2$ sublattice, the former seems to be associated with either intrinsic or extrinsic inhomogeneities in the material \cite{Xue01,Xue03}, see also \cite{Fel02,Fel03}. Recently, detailed susceptibility and magnetization measurements suggested that the higher of the two transitions may be due to the magnetic ordering of a small fraction of nanosized islands inside the crystal grains in which the Ru$^{4+}$ concentration is high. Alternatively, the transition could be due the presence of nanoparticles of a foreign minor extra Ru$^{4+}$ magnetic phase of Sr-Cu-Ru-O$_3$ in which Cu is distributed inhomogeneously in both the Ru and Sr sites \cite{Fel05}. Further, complex time dependent phenomena were found in dc or ac magnetic susceptibility measurements \cite{Ziv02,Xue03,Car03,Car04,Car05}, pointing perhaps to a spin glass behavior even at low temperatures, caused by the presence of oxygen vacancies in the RuO$_6$ octahedra  \cite{Car03,Car04,Car05}. In \cite{Ziv02} the interesting conclusion has been drawn that the complex behavior could be due to an antiparallel ordering of the in-plane magnetizations of adjacent RuO$_2$ layers. Then, similar to the case of the type I ordering proposed for Ru-1212 by Butera {\it et al.} \cite{But01} also Ru-1222 could exhibit an intrinsic spin valve effect in case there is an interlayer tunnel transport in this material. 

\subsection{Superconducting Properties}
As already mentioned, Ru-1212 becomes superconducting in the range 15-50\,K. The transition temperature for Ru-1222 is in the range 30-40\,K (cf. figure 2). For Ru-1212Gd Figure 8 shows an example for a resistively measured transition \cite{Ber99}.
\begin{figure}[h]
\centering
\epsfig{file=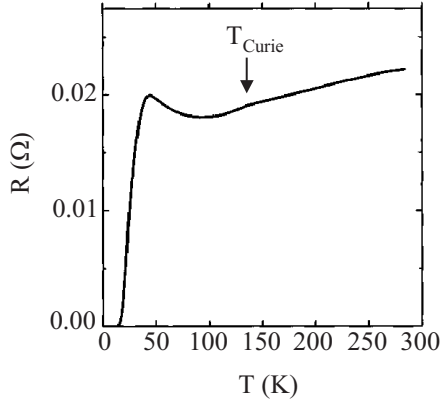}
\caption{Temperature dependence of the resistance of a polycrystalline Ru-1212Gd sample (from \cite{Ber99}).}
\end{figure}
Zero resistance is reached at $T_c = 16$\,K. The Curie temperature of this sample is about 133\,K. Note that near this temperature the resistance vs. temperature curve exhibits a small cusp. The transition to the superconducting state has been investigated with various methods, including measurements of specific heat \cite{Tal00,Che01} ensuring that the superconductivity observed is indeed a bulk effect coexisting with ferromagnetism. 

Similar to the cuprates also in the the ruthenocuprates the hole doping of the CuO$_2$ layers can be altered by substituting proper elements. For example, using a high pressure synthesis method, Klamut {\it et al.} \cite{Kla03,Kla01,Kla01b} investigated compounds of the form Ru$_{1-x}$Sr$_2$GdCu$_{2+x}$O$_{8+\delta}$ and RuSr$_2$Gd$_{1-y}$Ce$_y$Cu$_2$O$_8$. In the former compounds the Ru is partially substituted by Cu, leading to an additional doping of the CuO$_2$ layers with holes. For the latter compounds, by substituting trivalent Gd by Ce$^{4+}$ holes are extracted from CuO$_2$ layers. Figure 9 shows how the transition temperature $T_c$, as well as the magnetic phase transition temperatures ($T_N$ for RuSr$_2$Gd$_{1-y}$Ce$_y$Cu$_2$O$_8$, $T_m$ for Ru$_{1-x}$Sr$_2$GdCu$_{2+x}$O$_{8+\delta}$) evolve with doping.
\begin{figure}[h]
\centering
\epsfig{file=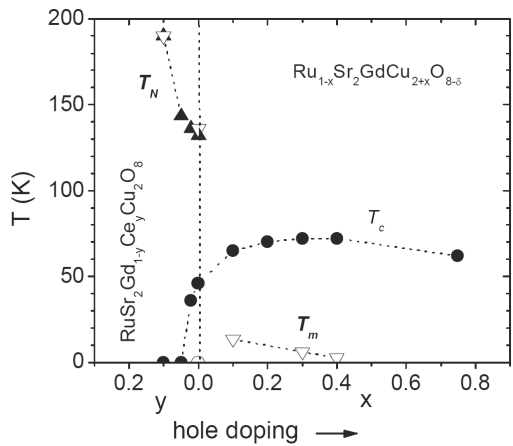}
\caption{Dependence of the superconducting transition temperature $T_c$ and the magnetic ordering temperatures ($T_m$ and $T_N$) on hole doping, as achieved for the compounds RuSr$_2$Gd$_{1-y}$Ce$_y$Cu$_2$O$_8$, and  Ru$_{1-x}$Sr$_2$GdCu$_{2+x}$O$_{8+\delta}$ (from \cite{Kla03}).} 
\end{figure}
For Ru$_{1-x}$Sr$_2$GdCu$_{2+x}$O$_{8+\delta}$ a maximum $T_c$ of 72\,K was achieved for $x = 0.4$, where $T_m$ is almost zero. For RuSr$_2$Gd$_{1-y}$Ce$_y$Cu$_2$O$_8$ $T_c$ strongly decreases with increasing $y$ while $T_N$ increases.
 
For Ru-1212Gd also the effect of pressure on the superconducting and magnetic transition has been investigated \cite{Lor03} and analyzed \cite{Cit05}. For the superconducting transition one finds $dT_c/dp \approx 1$\,K/GPa while for the magnetic transition a value of about 6.7\,K/GPa is obtained. The relatively small value $dT_c/dp$ is interpreted as a competition of ferromagnetic and superconducting phases: a stronger enhancement of the magnetic phase results in a reduced pressure effect on $T_c$ as compared to underdoped high-$T_c$ compounds \cite{Cit05}. 

Also first attempts have been made to perform point contact spectroscopy on Ru-1212Gd \cite{Giu03,Umm03,Pia05}.  It turns out that the tunneling curves exhibit zero bias conductance peaks (ZBCPs), as they are well known for the cuprates, see e. g. \cite{Tan00}. Although there are several possible origins of ZBCPs a likely one is that the ZPCPs observed are due to the formation of Andreev bound states. Such states are expected when the superconducting order parameter of Ru-1212 has nodes, e. g. in the case of $d$-wave symmetry. Further analysis of the conductance spectra yielded estimates of the superconducting energy gap of about 2.7\,meV in \cite{Pia05} and about 6\,mV in \cite{Umm03}, corresponding to $2\Delta/k_BT_c$ values of 2-4.

Before turning to further properties of the superconducting state it should be again noted that investigations have been performed on polycrystalline samples where intragrain and intergrain properties are superposed. For the high-$T_c$ cuprates it is well known that under such conditions most grains are coupled by weak Josephson currents forming a 3D Josephson network. A similar situation can be expected for the ruthenocuprate sinters and is indeed found. For example, from transport experiments on Ru-1222Gd the maximum supercurrent $I_c$ was found to be only of order 20\,A/cm$^2$ at 5\,K. In external fields as small as 10\,Oe $I_c$ is further suppressed by an order of magnitude, as it could be expected for intergrain Josephson junctions \cite{Fel03b}. Regarding the resistively measured transition a characteristic feature of Josephson networks is that, while the transition strongly broadens with increasing magnetic field the onset of the superconducting transition (given by the temperature where the grains become superconducting) shifts only slightly. This effect has been seen in transport experiments both on Ru-1212 and Ru-1222 \cite{Gar04,Gar03,Att04,Cim05}. A similar broadening occurs also in susceptibility measurements. Also many other measurements like microwave absorption \cite{Poz01} showed characteristic features of a Josephson network.

Regarding intragrain properties, the shift with magnetic field of $T_c$,onset in resistive or inductive measurements can be used to estimate the upper critical field $B_{c2}$.  It turns out that, as for the high-$T_c$ cuprates, also for the ruthenocuprates $B_{c2}$ seems to be quite high. For Ru-1212 zero temperature values between 28\,T and 80\,T have been extrapolated \cite{Tal00,Att04,Cim05}, corresponding to coherence lengths between 33 and 20\,\AA. Note that these numbers are spatial averages.
 
From magnetization and susceptibility measurements the (averaged) intragrain London penetration depth $\lambda_L$ ($T=0$) can be estimated. For Ru-1212Gd a value around  0.4-0.5\,$\mu$m has been found \cite{Ber00}, for Ru-1212Eu a value of about 1\,$\mu$m was obtained \cite{Lor02} and for Ru-1222 $\lambda_L$ ($T=0$) is about 2\,$\mu$m \cite{Xue02}. These numbers are indeed quite large and either correspond to very low superconducting condensate densities or else may point to inhomogeneties inside the grains. Such inhomogeneities could arise because superconducting regions in the grains are phase separated on a nanoscale, establishing macroscopic phase coherence via the Josephson effect. This possibility has been investigated intensively by the Texas group using various compounds \cite{Lor02,Xue02,Xue00,Chu00}. The authors conclude that the ruthenocuprates are indeed inhomogeneous. The same conclusion has been drawn in \cite{Gar03} for Ru-1212 and Ru-1222. On the other hand, these intragrain weak links could well be associated with the $c$-axis Josephson coupling, seen in Ru-1212Gd single crystals (cf. section 3).
 
Regarding intragrain critical current densities, from magnetization studies of Ru-1222Y a value of of some kA/cm$^2$ at low temperatures has been given \cite{Fel03}. Also this number is at least one or two orders of magnitude lower than the corresponding values in high-$T_c$ cuprates.

We next turn to the possible interplay between superconductivity and magnetism. While the superconducting transition of the CuO$_2$ bilayers is likely to have no noticeable effect on the magnetic ordering of the Ru ions, the superconducting state may well be affected by the magnetism of the Ru layers. 

The first question is to what extent the Ru magnetic moments can break Cooper pairs in the CuO$_2$ layers. If the exchange splitting in the CuO$_2$ layers caused by the magnetic moments of the Ru$^{5+}$ and Ru$^{4+}$ ions were on the order of the superconducting energy gap one might expect a FFLO state or even spin triplet superconductivity, as already mentioned in the introduction. Indeed, early estimates yielded values of the exchange splitting on the order of 20-50\,meV \cite{Pic99,McC99}. On the other hand, detailed band structure calculations by Nakamura {\it et al.} \cite{Nak01} that took structural distortions into account, yielded an exchange splitting between zero and 10\,meV, making a severe pair breaking effect in the superconducting condensate less likely. Also the Zeeman splitting due to dipolar fields caused by the volume averaged magnetization of the material is on the order of $\mu$eV and thus cannot break Cooper pairs \cite{Pic99}. In this case the interactions between the superconducting and the magnetic systems are more subtle, although some competition between the superconducting and magnetic transition temperatures seem to exist at least for Ru-1222 \cite{Awa04}. 

One example is the occurrence of a spontaneous vortex phase (SVP) caused by the weak ferromagnetic component of the canted Ru lattice. Such a state was proposed in the context of Rhodium borides (e. g. ErRh$_4$B$_4$) \cite{Gre81}. Basically, due to the presence of ferromagnetic order, in a temperature range $T_{Ms} < T < T_c$  vortices penetrate the material even without an applied magnetic field. 
\begin{figure}[h]
\centering
\epsfig{width=6cm,file=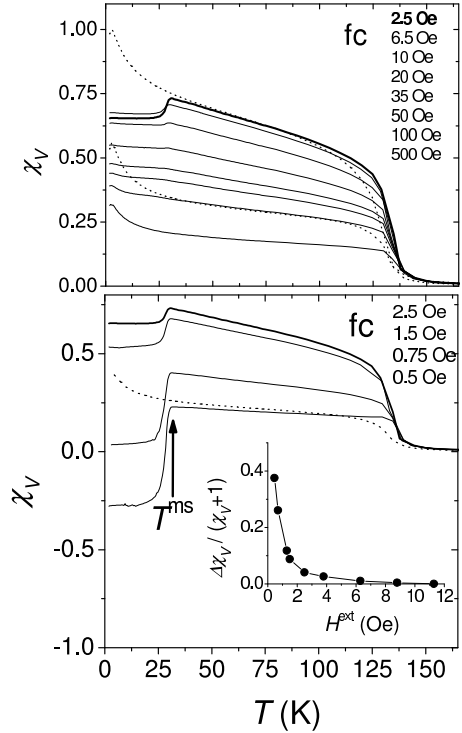}
\caption{Field cooled (fc) volume susceptibility $\chi_V$ vs. temperature of a pure Ru-1212Gd sample (solid lines) and a Zn substituted non-superconducting sample (dotted lines) for fields between 0.5\,Oe and 2.5\,Oe (lower figure) and between 2.5\,Oe and 500\,Oe (upper figure). Inset shows field depencence of the size of the diamagnetic shift $[\chi_V(T \rightarrow 0)- \chi_V(T_{ms})]/[ \chi_V(T_{ms})+1]$, yielding an estimate of the volume fraction of the Meissner phase. $T_{ms}$ denotes the temperature where the Meissner effect sets in (from \cite{Ber00}).}
\end{figure}
Only below $T_{Ms}$ the Meissner state establishes. Evidences for an SVP have been given e. g. in \cite{Son98,Fel00b,Ber00,Tok01}, see also \cite{Lev04}.  While in \cite{Son98,Fel00b} Eu$_{1.4}$Ce$_{0.6}$RuSr$_2$Cu$_2$O$_{10}$ was investigated, in \cite{Ber00} and \cite{Tok01} the focus was on, respectively, Ru-1212Gd and Ru-1212Y. In \cite{Tok01} the suggestion of a SVP was given on the basis of NMR data, as obtained from the Cu and Ru sites. In \cite{Ber00} for a sample with $T_c \approx 45$\,K dc magnetization measurements revealed the onset of a large diamagnetic signal corresponding to a bulk Meissner phase below $T_{ms} \approx 30$\,K, cf. figure 10. In the regime $T_{ms} < T < T_c$ unique thermal hysteresis effects were observed pointing to the SVP. However, it should be noted here that, in case the ruthenocuprates exhibit phase separation on a nanoscale, a Meissner phase would also not occur until macroscopic phase coherence between the nanograins is established. 

The second example which we address here in the context of polycrystalline samples and in section 3 in the context of single crystals is the prediction of $\pi$ phases, where the superconducting order parameter changes sign between adjacent superconducting layers. As already mentioned in the introductory section, if adjacent CuO$_2$ double layers are Josephson coupled, the $\pi$ state would correspond to an interlayer Josephson current-phase relation involving an additional phase factor of $\pi$. For Ru-1212Gd early measurements of the optical conductivity revealed no superconductivity-related features in the far-infrared response pointing to a rather low $c$-axis plasma frequency possibly associated with interlayer Josephson coupling \cite{Lit00}. Indeed, later on, via far-infrared spectroscopy Shibata observed a resonance at wavenumbers of about 8.5\,cm$^{-1}$ strongly indicative of Josephson plasma oscillations, as observed e. g. in BSCCO \cite{Shi02,Shi03}. However, the frequency of the resonance, for Josephson plasma oscillations being proportional to the square root of the maximum Josephson current density $j_c$, evolved monotonously with temperature. By contrast, if there were a transition between a conventional 0 state and a $\pi$ state at some temperature, a zero or at least a minimum of $j_c$ should occur at the $0-\pi$ transition. Shibatas measurements thus indicate that such a transition does not occur.  



\section{Ru-1212Gd single crystals: Magnetization and $c$-axis transport studies}
\label{singlecrystals}
In this section we focus to some detail on results obtained for Ru-1212Gd single crystals. Preparation conditions have been briefly addressed in the previous section, cf. figure 5. 

\subsection{Magnetization measurements}
For the magnetization measurements \cite{Nac_u} discussed here a relatively large crystal with dimensions $200 \times 200 \times 100\mu$m$^3$ was selected to perform SQUID magnetization measurements. Figure 11(a) shows results for magnetization vs. temperature (field cooled curves), as measured in four magnetic fields between 5\,Oe and 50\,Oe and for field orientations parallel and perpendicular to the layers. 
\begin{figure}[h]
\centering
\epsfig{file=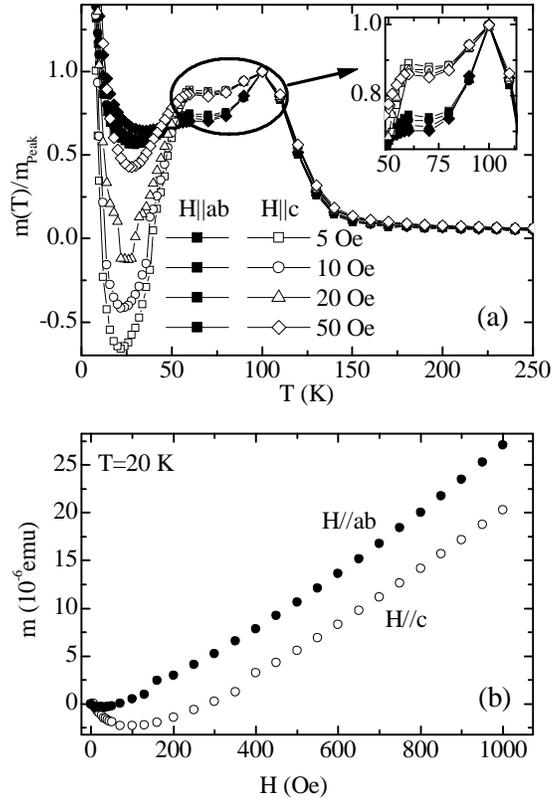}
\caption{Field cooled magnetization vs. temperature (a) and zero field cooled magnetization vs. applied field (b) for a Ru-1212Gd single crystal. Fields are either applied parallel ($H\|ab$) to the layers or perpendicular to them ($H\|c$). Magnetization curves in (a) are normalized to their value at the peak near 100\,K. Inset in (a) shows magnetization curves on an enlarged scale. }
\end{figure}
The magnetization is normalized to its value at the peak near 100\,K. Absolute values at the peak were e. g. at 50\,Oe $1.5 \times 10^{-6}$\, emu/mol for $H\|ab$ and $1 \times 10^{-6}$\,emu/mol for $H\|c$. Between room temperature and 100\,K all curves are on top of each other. Below about 140\,K (the onset of magnetic ordering) the magnetization strongly increases and then drops below 100\,K, indicating that antiferromagnetic ordering of the Ru sublattice sets in. The magnetization drops somewhat stronger for fields parallel than perpendicular to the layers. Although the effect is not pronounced enough to draw a strong conclusion, this may indicate that the Ru moments are oriented in-plane rather than out of plane. This result would contradict the results of neutron diffraction measurements but could be explained in the framework of a G-type or Type-I antiferromagnet with the easy axis in-plane. The difference to the neutron diffraction measurements may be related to the fact that the crystals are likely to have excess Cu in the RuO$_2$ layers (and thus the magnetic ordering may be different than in the Ru-1212 sinters). The excess Cu would also explain the relatively low value of the magnetic ordering temperature and the relatively high $T_c$.
 
Below the superconducting transition near 50\,K, for fields perpendicular to the layers the magnetization drops strongly, reaching negative values for $H<10$\,Oe. By contrast, the drop in magnetization is much less pronounced when the field is applied parallel to the layers. Such a behavior is indeed what can be expected for a Josephson coupled layer structure, as modelled in figure 1(c). While for the parallel orientation the magnetic field easily penetrates the crystal for the perpendicular configuration strong in-plane screening currents are excited, leading to the diamagnetic response. Finally, at temperatures below 20\,K the magnetization strongly increases due to the Gd magnetic moments. Figure 11(b) shows zero-field-cooled magnetization curves at 20\,K, with the magnetic field applied either parallel or perpendicular to the layers. Again, the diamagnetic response is seen only for $H\|c$ consistent with the above picture.

We next turn to interlayer transport experiments. In the context of intrinsic Josephson junctions in high-$T_c$ cuprates and intrinsic spin valves in layered manganites, several methods have turned out to be suitable to perform measurements on very small sized specimens. In early works single crystals with diameters some 10\,$\mu$m were clamped between two metallic contact rods, and measurements were performed in a two-terminal configuration. Later on combinations of e-beam and photolithography were used to pattern mesa structures some $\mu$m in diameter with thicknesses in the nm range on top of the crystals \cite{Kle92,Yur00,Nac01}. For the Ru-1212Gd crystals used, it turned out that the mesa technology produced too large contact resistance between the crystal surface and the contacting Au layers to be useful. 
\begin{figure}[h]
\centering
\epsfig{file=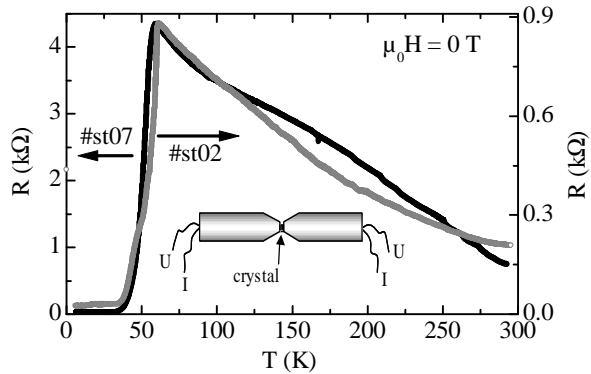}
\caption{Out-of-plane resistance of two Ru-1212Gd single crystals, as measured in the two-terminal configuration shown in the inset.}
\end{figure}
Crystals thus have been measured in a two terminal configuration, as indicated in the inset of figure 12 \cite{Nac04}. For the measurement single crystals of typical in-plane sizes of 50-100\,$\mu$m and thicknesses of 15-40\,$\mu$m have been clamped between two contact rods such that the $c$-axis of the crystals was perpendicular to the contacting area. Figure 12 shows the temperature dependence of the out-of-plane resistance for two crystals that have been grown in the same batch. For crystal st07 the midpoint of the resistive transition is at 51\,K, with a transition width of 10\,K. Crystal st02 has a slightly higher $T_c$ of 54\,K. Here, the transition shows a footlike structure below 50\,K. The residual resistance at low temperatures, due to the contact resistance between the contacting Au layer and the crystal, amounts to only a few per cent of the total resistance and is thus not a major concern. For both samples, $R(T)$ exhibits a maximum near 60\,K. From the crystal sizes ($75 \times 75 \times 20\,\mu$m$^3$ and $120 \times 120 \times 35\,\mu$m$^3$, respectively) at this maximum one estimates a resistivity $\rho_c \approx 25\,\Omega$cm for st02 and $170\,\Omega$cm for st07. Due to irregularities in the crystal shape and the possibility that the contacting gold layer has shunted part of the crystal side walls these numbers should however be taken with an error bar of at least 50\%. From the differences between the two samples we see that the crystals are certainly not perfectly homogeneous. On the other hand, the values of $\rho_c$ obtained here are of the same order as the out-of-plane resistances of BSCCO or LSMO, giving a first indication that the transport current indeed flows out of plane, since any shorts caused by currents flowing in-plane would drastically decrease the resistivity. For comparison, the resistivity measured for polycrystalline samples is in the m$\Omega$cm range, i. e. drastically lower. 

\begin{figure}[h]
\centering
\epsfig{file=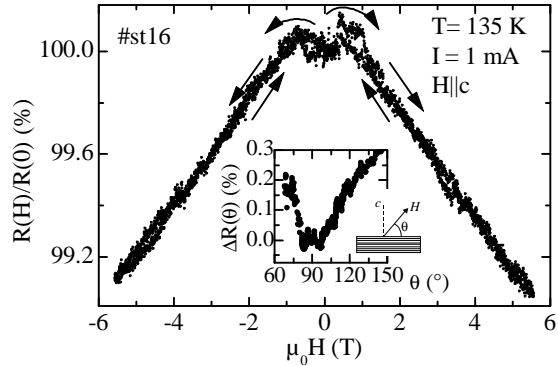}
\caption{Out-of-plane magnetoresistance of a Ru-1212Gd single crystal measured near the magnetic ordering temperature of about 135\,K. Magnetic field is applied perpendicular to the layers. Inset shows angle dependent magnetoresistance at $T = 135$\,K for a field of 6\,T.} 
\end{figure}
Figure 13 shows the out-of-plane magnetoresistance of single crystal st16, as measured near the magnetic ordering temperature of 135\,K for magnetic fields applied perpendicular to the layers. The magnetoresistance is positive for fields below 90\,mT but becomes negative for higher fields. Above 1\,T it decreases about linearly, with a slope $dR/dH \approx - 0.2\%$ per Tesla. A similar behavior was also observed for Ru-1212 and Ru-2212 polycrystalline samples \cite{McC99,Che01}, where the negative magnetoresistance was attributed to interactions between the charge carriers (flowing in-plane through either the CuO$_2$ or the RuO$_2$ planes) and the localized Ru spins becoming more ordered in external fields. As we will see below at least the interlayer supercurrents are of the tunneling type. The negative $c$-axis magnetoresistance shown in figure 13 thus may require other explanations based on interlayer tunneling processes. Further, the inset of figure 13 shows the angle dependence of the out-of-plane magnetoresistance, as measured at 135\,K in a field of 6\,T. The variation in $R(\theta)$ amounts to only 0.2\% but shows a clear minimum when the magnetic field approaches $\theta=90^\circ$. We note here that a similar high-field magnetoresistance was obtained for LSMO single crystals \cite{Nac01}. The effect was, however much stronger there and it was in addition accompanied with a low-field switching effect corresponding to the spin-valve effect mentioned in the introduction.
 
Let us now turn to data obtained in the superconducting state. Figure 14(a) shows a current voltage characteristic of a Ru-1212Gd single crystal at 4.2\,K. 
\begin{figure}[h]
\centering
\epsfig{file=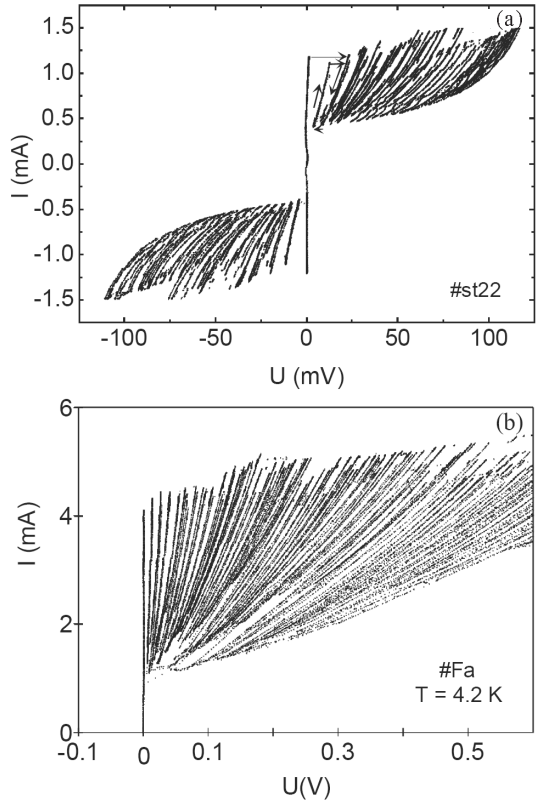}
\caption{Out-of-plane current-voltage characteristic of a Ru-1212Gd single crystal at 4.2\,K (a) in comparison with a BSCCO single crystal (b). Both crystals were measured in a two-terminal configuration. The resulting contact resistance has been subtracted in the graphs. The multiple hystereses have been traced out by repeatedly ramping the bias current up and down, as indicated in (a) for the first two resistive branches.}
\end{figure}
The contact resistance has been subtracted from the data. At a current of about 1.2\,mA a voltage jump to a resistive state occurs. Lowering the current in this resistive state the voltage decreases continuously down to about 0.5\,mA where another voltage jump occurs. By repeatedly increasing and decreasing the bias current the multiple branched structure can be traced out. The behavior seen here is well known from intrinsic Josephson junction stacks in BSCCO, where the multiple hysteresis arises, because the junctions in the stack can be switched from the superconducting state to the resistive state one by one. Figure 14(b) gives an example. The BSCCO single crystal investigated here has been measured in a similar sample holder as used for the Ru-1212Gd single crystals. Its thickness was 3\,$\mu$m, corresponding to a stack of about 2000 intrinsic junctions, much more than have been switched to the resistive state in the figure (in view of the model shown in figure 1(c) one expects an intrinsic junction per 1.2\,nm and thus well above 10000 junctions in the Ru-1212Gd single crystals used here). The branch structure of the Ru-1212Gd crystal looks less regular than the one of the BSCCO crystal indicating that the Ru-1212Gd crystal is less homogeneous. However, to be fair, also many high-$T_c$ intrinsic junction stacks often exhibit current voltage characteristics that are less regular than the one shown here (see e. g. \cite{Kle92,Kle94}). Nonetheless one needs to confirm that the branching observed here indeed comes from a junction stack and is not due to e. g. Josephson junctions between some inhomogeneous superconducting regions in the crystals. 

The motion of Josephson flux quanta (fluxons) along the barrier layers can be used to investigate the junction orientation. When a large enough magnetic field is applied parallel to the barrier layers and the junction length perpendicular to the applied field is well above the Josephson penetration depth (typically around a few micrometers) fluxons form and can be driven along the junction due to the Lorentz force created by the bias current. The motion of fluxons starts as soon as pinning forces arising e. g. from the edges of the junctions or some inhomogeneities in the interior of the junctions are overcome. When the magnetic field is tilted away from the junction barrier layer by an angle $\theta$ (cf. figure 15, inset) the field penetrates the superconducting electrodes in the form of Abrikosov vortices (or pancake vortices in the case of atomically thin superconducting layers) as soon as $\theta$ becomes larger than some critical angle $\theta_c$, which is essentially determined by the perpendicular field component increasing over the lower critical field of the superconducting layers.
\begin{figure}[h]
\centering
\epsfig{file=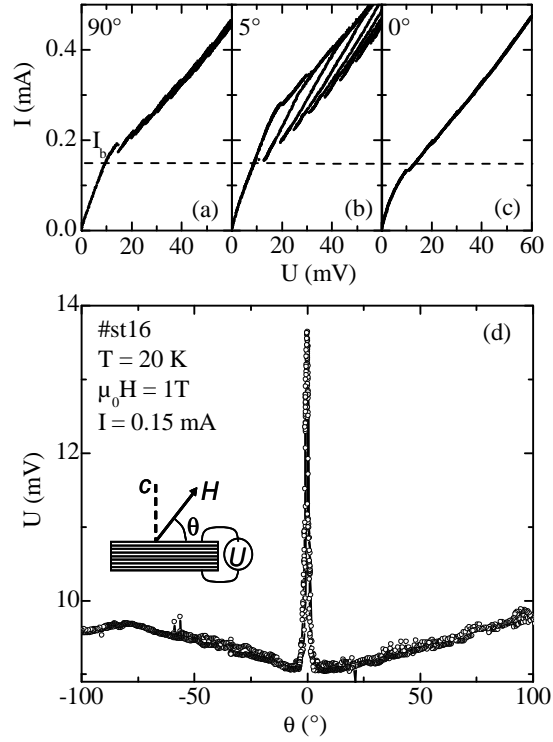}
\caption{Out-of-plane current voltage characteristics of a Ru-1212Gd single crystal, as measured at various angles $\theta$ between the applied magnetic field and the crystallographic $c$-axis: (a) $\theta=90^\circ$, (b) $\theta=5^\circ$ and (c) $\theta=0^\circ$. Graph (d) shows the voltage drop across the crystal at a fixed bias current of 0.15\,mA as a function of $\theta$. Dashed line in (a) to (c) also indicates this bias current.}
\end{figure}
The tilted flux line then consists of a Josephson fluxon string terminated by Abrikosov vortices. The latter strongly pin the Josephson fluxon string. For a stack of intrinsic Josephson junctions the above effect is known as the lock-in transition and is routinely used to align BSCCO intrinsic Josephson junctions in external magnetic fields \cite{Iri01}. Usually a not too small temperature is chosen for the measurements (typically 50-60\,K for BSCCO) to avoid hysteresis with respect to $\theta$ (pancake vortices are only weakly pinned within the CuO$_2$ layers at these temperatures). Figure 15 shows a corresponding measurement for a Ru-1212Gd single crystal. Figures 15(a) to (c) show current voltage characteristics measured at different misorientation angles. Measurements were done in a field of 1\,T at a temperature of 20\,K. For $\theta = 90^\circ$, i. e. with the field applied perpendicular to the layers the multiple branching has almost disappeared due to the presence of a large number of pancake vortices. Lowering $\theta$ the multiple branching becomes again apparent, as can be seen in figure 15(c) for a misorientation angle of 5$^\circ$. At perfect alignment ($\theta = 0^\circ$) the branches are again suppressed and the current voltage characteristic is dominated by Josephson vortex flow, cf. figure 15(c). When the bias current is fixed (at $I_b = 0.15$\,mA for the example discussed here) and the voltage across the crystal is measured as a function of $\theta$ while rotating the sample, one finds the curve shown in figure 15(d). There is a sharp peak around $\theta = 0^\circ$ fully consistent with the well known case of BSCCO single crystals. No other peaks are observable. By contrast, had the crystal consisted of a 3D network of Josephson junctions (e. g. formed between superconducting grains inside the sample) having orientations distributed randomly, peaks would be observable at many angles. Figure 15(d) thus shows that the barrier layers of the junctions measured are all in parallel and indeed parallel to the $ab$-direction of the crystal. The crystal thus indeed forms a well defined stack of intrinsic Josephson junctions.

The next question that must be addressed is whether or not intrinsic $\pi$ junctions are formed. As long as the system is in either the 0 state or the $\pi$ state the question is indeed hard to answer. On the other hand, a transition between a 0 state and a $\pi$ state would result in a vanishing interlayer critical current $I_c$ at the $0-\pi$ transition. The effect has been clearly seen in Nb/Cu-Ni/Nb junctions \cite{Rya01} where $I_c$ vs. temperature went through a zero. In terms of the ruthenocuprates Houzet {\it et al.} \cite{Hou01} have calculated a temperature-exchange field phase diagram  showing that transitions between a 0 state and a $\pi$ state can be expected both in terms of temperature and exchange magnetic field. Figure 16 shows the phase diagram. 
\begin{figure}[h]
\centering
\epsfig{file=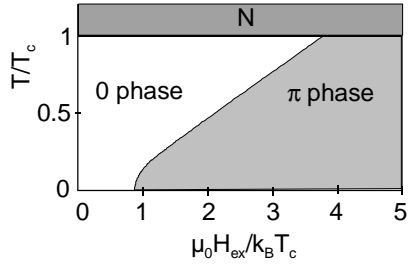}
\caption{Predicted temperature - exchange field diagram showing transitions between the 0 state, the $\pi$ state and the normal conducting (N) state. Temperature is given in units of the superconducting transition temperature, exchange field $H_{ex}$ in units of $k_BT_c/\mu_0$ (after \cite{Hou01}).}
\end{figure}
Note that, if the temperature is varied, $0-\pi$ transitions can be expected in some finite range of the normalized exchange energy $\mu_0H_{ex}/k_BT_c$. In case of $s$-wave pairing symmetry one obtains $0.87 < \mu_0H_{ex}/k_BT_c < 3.77$ while for $d$-wave pairing one gets  $0.6 < \mu_0H_{ex}/k_BT_c < 3.77$. One finds $\mu_0H_{ex}  = J^{ab}S_{\rm\it eff}$, where $J^{ab}$ is the in-plane exchange integral and $S_{\rm\it eff}$ is an effective spontaneous spin given by the in-plane magnetisation normalized to its saturation value. For zero applied field $S_{\rm\it eff} \approx 0.1$. Estimating  $J^{ab}/k_B = 100-200$\,K (band calculations \cite{Pic99} gave an estimate of 107\,K) one finds  $\mu_0H_{ex}/k_B$ to be of order 10-20\,K. In that case a nonmonotonous $I_c$ vs. $T$ curve should be expected, having a zero or at least a minimum at the $0-\pi$ transition. 

\begin{figure}[h]
\centering
\epsfig{file=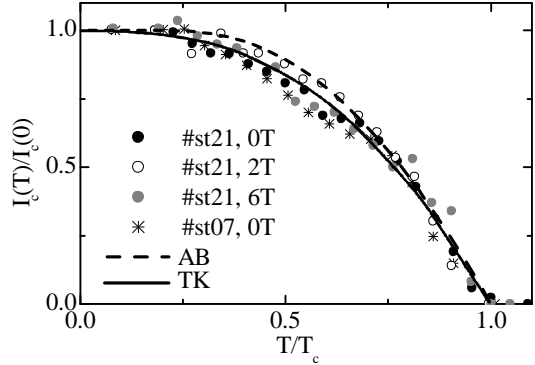}
\caption{Temperature dependence of the normalized out-of-plane critical current of two Ru-1212Gd single crystals at different magnetic fields. Dashed line is the Ambegaokar-Baratoff curve for conventional SIS tunnel junctions, solid line corresponds to $I_c$ vs. $T$, as calculated by Tanaka and Kashiwaya \cite{Tan97} for a $d$-wave superconducting order parameter.}
\end{figure}
Figure 17 shows $I_c$ vs. $T$ for two Ru-1212Gd single crystals, st07 and st21. For st07 data are for 0\,T, for st21 data are for magnetic fields of  0, 2\,T and 4\,T, applied parallel to the layers. In all cases the first voltage jump in the current voltage characteristic has been used to determine the critical current. As can be seen, the data are a smooth function of temperature, with no indication of a $0-\pi$ transition. The data are also compared to the Ambegaokar-Baratoff (AB) result for conventional ($s$-wave) SIS tunnel junctions (dashed line) and to the result by Tanaka and Kashiwaya (TK) \cite{Tan97} for interlayer $d$-wave junctions (solid line). As can be seen, TK slightly fits better than AB, although the difference between the theoretical curves is not significant enough to take this curve as evidence for $d$-wave symmetry of the superconducting order parameter of Ru-1212Gd.

Applying an in-plane external field to the crystals increases the in-plane magnetization and thus $S_{\rm\it eff}$. For the crystals used, the magnetization due to the Ru sublattice saturated at fields of order 1-3\,T. Using the above estimate for $J^{ab}$, $\mu_0H_{ex}/k_B$ should thus increase up to the 100-200\,K range and the ratio $\mu_0H_{ex}/k_BT_c$ should increase to values of order 3-4. Figure 15 would thus result in a field driven $0-\pi$ transition. The $I_c$ vs. $T$ curves in figure 17 measured in fields of 2\,T and 6\,T indicate already that such a transition may not happen in the magnetic field range investigated. To investigate the possibility of a field driven $0-\pi$ transition more exactly it is possible to use the lock-in peak observed when rotating the crystal into parallel alignment with field (cf. figure 15). Figure 18 shows $U$ vs. $\theta$ for a Ru-1212Gd single crystal as measured at various magnetic fields up to 4\,T at a temperature of 25\,K. 
\begin{figure}[h]
\centering
\epsfig{file=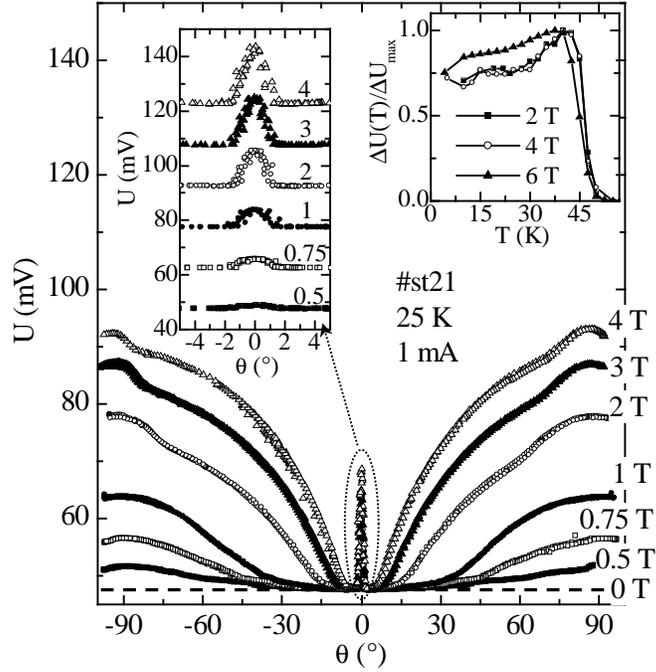}
\caption{Dependence of the voltage $U$ across a Ru-1212Gd single crystal on the angle $\theta$ between the applied field and the crystallographic $c$-axis. Curves are for 7 different values of magnetic field at $T = 24$\,K. Left inset shows an enlargement of the peak near $\theta=0^\circ$. Right inset shows the peak height vs. temperature for three different values of the applied field (from \cite{Nac04}).}
\end{figure}
As can be seen the Josephson flux flow peak near $\theta = 0^\circ$ increases monotonically with increasing field. In the absence of a $0-\pi$ transition such an increase is expected because the number of fluxons in the crystals increases proportional to the applied field. Near a $0-\pi$ transition the interlayer Josephson coupling should have vanished resulting in a disappearance of the Josephson flux flow peak. The inset shows the height of the peak as a function of temperature for three values of applied field. With increasing temperature the peak height weakly increases initially. Close to $T_c$ the peak height drops towards zero, an effect which is understandable, because for $T \rightarrow T_c$ the interlayer coupling vanishes. For lower temperatures, again no unconventional behavior is observed. It thus seems that a $0-\pi$ transition is absent in the Ru-1212Gd single crystals investigated. Figure 19 summarizes the values of temperature and magnetic field investigated. 
\begin{figure}[h]
\centering
\epsfig{file=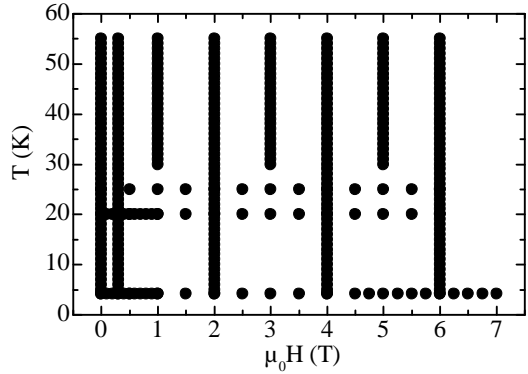}
\caption{Summary of data points to investigate the possibility of a 0-$\pi$ transition in Ru-1212Gd single crystals. Each point corresponds to a measurement of either the critical current or the Josephson flux flow peak in $U$ vs. $\theta$.}
\end{figure}
The temperature regime investigated corresponds to $0.075 < T/T_c < 1$. The magnetic field range, assuming the above estimates for $S_{\rm\it eff}$ and $J^{ab}$ would correspond to roughly $(0.2-0.4 )< \mu_0H_{ex}/k_BT_c < (2-4)$. Under these conditions the diagram shown in figure 16 would have been almost fully covered, although certainly only with a discrete number of data points. Nonetheless it seems unlikely that the $0-\pi$ transition has been missed in this regime. 

Note that the absence of a $0-\pi$ transition leaves open the question whether a $\pi$ state or a 0 state is realized. If the ferromagnetic layer is too weak even for $S_{\rm\it eff} = 1$ the crystal would always be in the 0 state. Another option is that the crystals were always in the $\pi$ state, i. e. $H_{ex}$ is too strong to observe a $0-\pi$ transition. The proof of such a possibility would require phase sensitive measurements of the superconducting order parameter in adjacent CuO$_2$ double layers which has not been done yet. Finally, a more complex state may be realized not requiring a $0-\pi$ transition, e. g. due to the formation of magnetic domains or due to a FFLO state with a superconducting order parameter not changing sign between adjacent superconducting layers. These possibilities certainly should motivate further investigations.

\section{Conclusions}
\label{conclusions}
As we have seen there is good evidence that in the ruthenocuprates superconductivity and magnetism indeed coexist, although the material quality is certainly far from being perfect. Also, many questions regarding the exact type of magnetic as well as superconducting ordering are still open. Most investigations so far have been performed on polycrystalline samples such that detailed information on anisotropy issues is not available yet. For Ru-1212Gd single crystals magnetization and interlayer transport experiments were possible, showing that adjacent CuO$_2$ double layers are Josephson coupled. More detailed investigations can be done as soon as larger crystals and good quality thin films become available. This is likely to happen in the near future.

\section*{Acknowledgments}
Financial support by the Deutsche Forschungsgemeinschaft, the Landesforschungs\-schwerpunktsprogramm Baden-W\"urttemberg and the ESF programs "Pi-Shift" and "Vortex" is gratefully acknowledged.


\begin{thebibliography}{00}

\bibitem{Ima98} M. Imada, A. Fujimori, Y. Tokura,
Rev. Mod. Phys. {\bf 70} (1998) 1039.
\bibitem{Kle92} R. Kleiner, F. Steinmeyer, G. Kunkel, P. M\"uller, Phys.Rev. Lett. {\bf 68} (1992) 2394.
\bibitem{Yur00} For a recent review, see A. Yurgens, Supercond. Sci. Technol. {\bf 13} (2000) R85.
\bibitem{Kim96} T. Kimura, Y. Tomioka, H. Kuwahara, A. Asamitsu, M. Tamura, Y. Tokura, Science {\bf 274} (1996) 1698.
\bibitem{Wel99} U. Welp, A. Berger D. J. Miller, V. K. Vlasko-Vlasov, K. E. Gray, J. F. Mitchell, Phys. Rev. Lett. {\bf 83} (1999) 4180.
\bibitem{Nac01} T. Nachtrab, S. Heim, M. M\"o\ss le, R. Kleiner, R. Koch, O. Waldmann, P. M\"uller, T. Kimura, Y. Tokura, Phys. Rev. B {\bf 65} (2001) 012410.
\bibitem{Fel97} I. Felner, U. Asaf, Y. Levi, O. Millo, Phys. Rev. B {\bf 55} (1997) R3374.
\bibitem{Son98} E. B. Sonin, I. Felner, Phys. Rev. B {\bf 57} (1998) R14000. 
\bibitem{Pic99} W. E. Pickett, R. Weht, A. B. Shick, Phys. Rev. Lett. {\bf 83} (1999) 3713.
\bibitem{Ber00} C. Bernhard, J. L. Tallon, E. Br\"ucher, R. K. Kremer, Phys. Rev. B {\bf 61} (2000) R14960.
\bibitem{Son02} E. B. Sonin, Phys. Rev. B {\bf 66} (2002) 100504(R).
\bibitem{Zhu00} J. X. Zhu, C. S. Ting, C. W. Chu, Phys. Rev. B {\bf 62} (2000) 11369.
\bibitem{Shi00} H. Shimahara, S. Hata, Phys. Rev. B {\bf 62} (2000) 14541.
\bibitem{Ful60} P. Fulde, R. A. Ferrell, Phys. Rev. A {\bf 135} (1964) 550. 
\bibitem{Lar65} A. I. Larkin, Yu. N. Ovchinnikov, Sov. Phys. JETP {\bf 20} (1965) 762.
\bibitem{Hou01} M. Houzet, A. Buzdin, M. Kuli\'c, Phys. Rev. B {\bf 64} (2001) 184501.
\bibitem{Pro99} V. Proki\'c, A. I. Buzdin, L. Dobrosavljevi\'c-Gruji\'c, Phys. Rev. B {\bf 59} (1999) 587.
\bibitem{Tal00} J. L. Tallon, J. W. Loram, G. V. M. Williams, C. Bernhard, Phys. Rev. B {\bf 61} (2000) R6471.
\bibitem{Mac03} A. P. Mackenzie, Y. Maeno, Rev. Mod. Phys. {\bf 75} (2003) 657-712.
\bibitem{Bau95} L. Bauernfeind, W. Widder, H. F. Braun, Physica C {\bf 254} (1995) 151.
\bibitem{Bau96} L. Bauernfeind, W. Widder, H. F. Braun, J. Low. Temp. Phys. {\bf 105} (1996) 1605.
\bibitem{Fel00} I. Felner, U. Asaf, Y. Levi, O. Millo, Physica C {\bf 334} (2000) 141.
\bibitem{Ber99} C. Bernhard, J. L. Tallon, C. Niedermayer, T. Blasius, A. Golnik, E. Br\"ucher, R. K. Kremer, D. R. Noakes, C. E. Stronach, E. J. Ansaldo, Phys. Rev. B {\bf 59} (1999) 14099.
\bibitem{Tal99} J. L. Tallon, C. Bernhard, M. Bowden, P. Gilberd, T. Stoto, D. Pringle, IEEE Trans. Appl. Supercond. {\bf 9} (1999) 1696.
\bibitem{Vas04} A. A. Vasiliev, M. Aindow, Z. H. Han, J. I. Budnik, W. A. Hines, P. W. Klamut, M. Maxwell, B. Dabrowski, Appl. Phys. Lett. {\bf 85} (2004) 3217.
\bibitem{She04} A. Shengelaya, R. Khasov, D. G. Eshchenko, I. Felner, U. Asaf, I. M. Savi\'c, H. Keller, and K. A. M\"uller, Phys. Rev. B {\bf 69} (2004), 024517.
\bibitem{Pap03} T. P. Papageorgiou, H. F. Braun, T. G\"orlach, M. Uhlarz, H. v. L\"ohneysen, Phys. Rev. B {\bf 68} (2003) 144518.
\bibitem{Pap02} T. P. Papageorgiou, H. F. Braun, and T. Hermannsd\"orfer, Phys. Rev. B {\bf 66} (2002) 104509.
\bibitem{Lin01} C. T. Lin, B. Liang, C. Ulrich, C. Bernhard, Physica C {\bf 364-365} (2001) 373.
\bibitem{Wat04} M. Watanabe, D. P. Hai, K. Kadowaki, Proceedings of the 4th International Symposium on "Intrinsic Josephson Effect and Plasma Oscillations in High-$T_c$ Superconductors", Tsukuba (2004).
\bibitem{Gom04} M. Gombos, A. Veccione, R. Clancio, D. Sisti, S. Uthayakumar, S. Pace, Physica C {\bf 408} (2004) 189.
\bibitem{Leb05} O. I. Lebedev, G. Van Tendeloo, G. Christiani, H.-U. Habermeier, A. T. Matveev, Phys. Rev. B {\bf 71} (2005), 134523.
\bibitem{Mat04} A. T. Matveev, G. Christiani, E. Sader, V. Damljanovic, H.-U. Habermeier, Physica C {\bf 417} (2004) 50.
\bibitem{Chm00} O. Chmaissem, J. D. Jorgensen, H. Shaked, P. Dollar, J. L. Tallon, Phys. Rev. B {\bf 61} (2000) 6401.
\bibitem{McL99} A. C. McLaughlin, W. Zhou, J. P. Atfield, A. N. Fitch, J. L. Tallon, Phys. Rev. B {\bf 60} (1999) 7512.
\bibitem{Nak01} K. Nakamura, K. T. Park, A. J. Freeman, Phys. Rev. B {\bf 63} (2001) 024507.
\bibitem{McL03} A. C. McLaughlin, J. P. Atfield, U. Asaf, I. Felner, Phys. Rev. B {\bf 68} (2003) 014503.
\bibitem{Liu01} R. S. Liu, L.-Y. Jang, H.-H. Hung, J. L. Tallon, Phys. Rev. B {\bf 63} (2001) 212507.
\bibitem{Tok01} Y. Tokunaga, H. Kotegawa, K. Ishida, Y. Kitaoka, H. Takagiwa, J. Akimitsu, Phys. Rev. Lett. {\bf 86} (2001) 5767.
\bibitem{Kum01} K. Kumagai, S. Takada, Y. Furukawa, Phys. Rev. B  {\bf 63} (2001) 180509.
\bibitem{Man02} P. Mandal, A. Hassen, J. Hemberger, A. Krimmel, A. Loidl, Phys. Rev. B {\bf 65} (2002) 144506.
\bibitem{Nak02} K. Nakamuram A. J. Freeman, Phys. Rev. B {\bf 66} (2002) R140405.
\bibitem{Ali04} A. A. Aligia, M. A. Gusm\"ao, Phys. Rev. B {\bf 70} (2004) 054403.
\bibitem{But01} A. Butera, A. Fainstein, E. Winkler, J. Tallon, Phys. Rev. B {\bf 63} (2001) 054442.
\bibitem{McC03} J. E. McCrone, J. L. Tallon, J. R. Cooper, A. C. MacLaughlin, J. P. Attfield, C. Bernhard, Phys. Rev. B {\bf 68} (2003) 064514.
\bibitem{Liu05} C.-J. Liu, C.S. Sheu, T.-W. Wu, L.-C. Huang, F. H. Hsu, H. D. Yang, G. V. M. Williams, C.-J. C. Liu, Phys. Rev. B {\bf 71} (2005) 014502.
\bibitem{Poz02} M. Po$\check{\rm z}$ek, A. Dul$\check{\rm c}$i\'c, D. Paar, A. Hamzi\'c, E. Tafra, G. V. M. Williams, S. Kr\"amer, Phys. Rev. B {\bf 65} (2002) 174514.
\bibitem{Mar02} M. De Marco, D. Coffey, J. Tallon, M. Haka, S. Torongian, J. Fridmann, Phys. Rev. B {\bf 65} (2002) 212506.
\bibitem{Sak03} H. Sakai, N. Osawa, K. Yoshimura, M. Fang, K. Kosuge, Phys. Rev. B {\bf 67} (2003) 184409.
\bibitem{Fel99} I. Felner, U. Asaf, C. Godart, E. Alleno, Physica B {\bf 259-261} (1999) 703.
\bibitem{Wil02} G. V. M. Williams, L.-Y. Jang, R. S. Liu, Phys. Rev. B {\bf 65} (2002) 064508.
\bibitem{Wil01} G. V. M. Williams, M. Ryan, Phys. Rev. B {\bf 64} (2001) 094515.
\bibitem{Hir02} Y. Hirai, I. $\check{\rm Z}$ivkovi\'c, B. H. Frazer, A. Reginelli, L. Perfetti, D. Ariosa, G. Margaritondo, M. Prester, D. Drobac, D. T. Jiang, Y. Hu, T. K. Sham, I. Felner, M. Pederson, M. Onellion, Phys. Rev. B {\bf 65} (2002) 054417.
\bibitem{Lyn00} J. W. Lynn, B. Keimer, C. Ulrich, C. Bernhard, J. L. Tallon, Phys. Rev. B {\bf 61} (2000) R14964.
\bibitem{Jor01} J. D. Jorgensen, O. Chmaissem, H. Shaked, S. Short, P. W. Klamut, B. Dabrowski, J. L. Tallon, Phys. Rev. B {\bf 63} (2001) 054440.
\bibitem{Tak01}  H. Takagiwa, J. Akimitsu, H. Kawono-Furukawa, H. Yoshizawa, J. Phys. Soc. Jap. {\bf 70} (2001) 333.
\bibitem{Weh99} R. Weht, A. Shick, W. E. Pickett, in {\it High Temperature Superconductivity}, edited by S. E. Barnes, J.Ashkenazi, J. L. Cohn, F. Zuo, AIP Conf. Proc. No. {\bf 483} (AIP, New York, 1999), 141-146.
\bibitem{Fai99} A. Fainstein, E. Winkler, A. Butera, J. Tallon, Phys. Rev. B {\bf 60} (1999) R12597.
\bibitem{Xue01} Y. Y. Xue, D. H. Cao, B. Lorentz, C. W. Chu, Phys. Rev. B {\bf 65} (2001) R020511.
\bibitem{Xue03} Y. Y. Xue, B. Lorenz, D. H. Cao, C. W. Chu, Phys. Rev. B {\bf 67} (2003), 184507.
\bibitem{Fel02} I. Felner, U. Asaf, E. Galstyan, Phys. Rev. B {\bf 66} (2002), 024503.
\bibitem{Fel03} I. Felner, E. Galstyan, Int. J. Mod. Phys. {\bf 17} (2003) 3617.
\bibitem{Fel05} I. Felner, E. Galstyan, I. Nowik, Phys. Rev. B {\bf 71} (2005) 064510.
\bibitem{Ziv02} I. $\check{\rm Z}$ivkovi\'c, Y. Hirai, B. H. Frazer, M. Prester, D. Dobrac, D. Ariosa, H. Berger, D. Pavuna, G. Margaritondo, I. Felner, M. Onellion, Phys. Rev. B {\bf 65} (2002), 144420.
\bibitem{Car03} C. A. Cardoso, F. M. Araujo-Moreira, V. P. S. Awana, E. Takayama-Muromachi, O. F. de Lima, H. Yamauchi, M. Karppinen, Phys. Rev. B {\bf 67} (2003) 020407(R).
\bibitem{Car04} C. A. Cardoso, F. M. Araujo-Moreira, V. P. S. Awana, H. Kishan, E. Takayama-Muromachi, O. F. de Lima, Physica C {\bf 405} (2004) 212.
\bibitem{Car05} C. A. Cardoso, A. J. C. Lanfredi, A. J. Chiquito, F. M. Araúro-Moreira, V. P. S. Awana, H. Kishan, R. L. de Almeida, O. F. de Lima, Phys. Rev. B {\bf 71} (2005) 134509.
\bibitem{Che01} X. H. Chen, Z. Sun, K. Q. Wang, S. Y. Li, M. Xiong, M. Yu, L. Z. Cao, Phys. Rev. B {\bf 63} (2001) 064506.
\bibitem{Kla03} P. W. Klamut, B. Dabrowski, S. M. Mini, M. Maxwell, J. Mais, I. Felner, U. Asaf, F. Ritter, A. Shengelaya, R. Khasanov, I. M. Savic, H. Keller, A. Wisniewski, R. Puzniak, I. M. Fita, C. Sulkowski, and M. Matusiak, Physica C {\bf 387} (2003) 33.
\bibitem{Kla01} P. W. Klamut, B. Dabrowski, S. Kolesnik, M. Maxwell, J. Mais, Phys. Rev. B {\bf 63} (2001) 224512.
\bibitem{Kla01b} P. W. Klamut, B. Dabrowski, J. Mais, M. Maxwell, Physica C {\bf 350} (2001) 24.
\bibitem{Lor03} B. Lorenz, R. L. Meng, Y. Y. Xue, C. W. Chu, Physica C {\bf 383} (2003) 337. 
\bibitem{Cit05} R. Citro, G. G. N. Anginella, M. Marinaro, R. Pucci, Phys. Rev. B {\bf 71} (2005) 134525.
\bibitem{Giu03} F. Giubileo, F. Bobba, M. Gombos, S. Uthayakumar, A. Veccione, A. I. Akimenko, A. M. Cucolo, Int. J. Mod. Phys. B {\bf 17} (2003) 3525.
\bibitem{Umm03} G. Ummarino, A. Calzolari, D. Daghero, R. Gonnelli, V. Stepanov, R. Masini, M. Cimberle, {\it Proceedings of the 6th EUCAS Conference (14-18 Sep. 2003, Sorrento, Italy)} cond-mat/0309553 (2003).
\bibitem{Pia05} S. Piano, F. Bobba, F. Giubileo, A. M. Cucolo, Int. J. Mod. Phys. B {\bf 19} (2005) 323.
\bibitem{Tan00} S. Kashiwaya, Y. Tanaka, Rep. Prog. Phys. {\bf 63} (2000) 1641.
\bibitem{Fel03b} I. Felner, E. Galstyan, B. Lorenz, D. Cao, Y. S. Wang, Y. Y. Xue, and C. W. Chi, Phys. Rev. B {\bf 67} (2003), 134506.
\bibitem{Gar04}  S. Garc\'ia, J. E. Musa, R. S. Freitas, L. Ghifelder, Phys. Rev. B {\bf 68} (2003) 144512.
\bibitem{Gar03} S. Garc\'ia, L. Ghivelder, Rhys. Rev. B {\bf 70} (2004) 052503. 
\bibitem{Att04} C. Attanasio, M. Salvato, R. Ciancio, M. Gombos, S. Pace, S. Uthayakumar, A. Veccione, Physica C {\bf 411} (2004) 126.
\bibitem{Cim05} M. R. Cimberle, M. Tropeano, M. Feretti, A. Martinelli, C. Artini, G. A. Costa, Supercond. Sci. Technol {\bf 18} (2005) 454.
\bibitem{Poz01} M. Po$\check{\rm z}$ek, A. Dul$\check{\rm c}$i\'c, D. Paar, G. V. M. Williams, S. Kr\"amer, Phys. Rev. B {\bf 64} (2001) 064508.
\bibitem{Lor02} B. Lorenz, Y. Y. Xue, R. L. Meng, C. W. Chu, Phys. Rev. B {\bf 65} (2002) 174503.
\bibitem{Xue02} Y. Y. Xue, B. Lorenz, A. Bailakov, D. H. Cao, Z. G. Li, C. W. Chu, Phys. Rev. B {\bf 66} (2002) 014503.
\bibitem{Xue00} Y. Y. Xue, S. Tsui, J. Cmaidalka, R. L. Meng, B. Lorenz, C. W. Chu, Physica C {\bf 341} (2000) 483.
\bibitem{Chu00} C. W. Chu, Y. Y. Xue, S. Tsui, J. Cmaidalka, A. K. Heilman, B. Lorenz, R. L. Meng, Physica C {\bf 335} (2000) 231.
\bibitem{McC99} J. E. McCrone, J. R. Cooper, J. L. Tallon, J. Low Temp. Phys. {\bf 117} (1999) 1199.
\bibitem{Awa04} V. P. S. Awana, M. A. Ansari, A. Gupta, R. B. Saxena, H. Kishan, D. Buddhikot, S. K. Malik, Phys. Rev. B {\bf 70} (2004) 104520.
\bibitem{Gre81} H. S. Greenside, E. I. Blount, C. M. Varma, Phys. Rev. Lett. {\bf 46} (1981) 49.
\bibitem{Fel00b} I. Felner, E. B. Sonin, T. Machi, N. Koshizuka, Physica C {\bf 341} (2000) 715.\bibitem{Lev04}  G. I. Leviev, M. I. Tsindlekht, E. B. Sonin, I. Felner, Phys. Rev. B {\bf 70} (2004) 212503.
\bibitem{Lit00} A. P. Litvinchuk, M. N. Iliev, Y.-Y. Xue, R. L. Meng, C. W. Chu, V. N. Popov, Phys. Rev. B {\bf 62} (2000) 9709.
\bibitem{Shi02} H. Shibata, Phys. Rev. B {\bf 65} (2002) 180507.
\bibitem{Shi03} H. Shibata, Physica C {\bf 388-389} (2003) 459.
\bibitem{Nac_u} T. Nachtrab, D. Koelle, R. Kleiner, C. Bernhard, C. T. Lin, R. Koch, P. M\"uller, unpublished.
\bibitem{Nac04} T. Nachtrab, D. Koelle, R. Kleiner, Ch. Bernhard, and C. T. Lin., Phys. Rev. Lett. {\bf 92} (2004) 11700.
\bibitem{Kle94} R. Kleiner, P. M\"uller, Phys. Rev. B {\bf 49} (1994) 1327. 
\bibitem{Iri01} A. Irie, S. Heim, S. Schromm, M. M\"o\ss le, T. Nachtrab, M. Godo, R. Kleiner, P. M\"uller, G. Oya, Phys. Rev. B {\bf 62} (2000) 6681.
\bibitem{Rya01} V. V. Ryazanov, V. A. Oboznov, A. Y. Rusanov, A. V. Veretenninov, A. A. Golubov, J. Aarts, Phys. Rev. Lett. {\bf 86} (2001) 2427.
\bibitem{Tan97} Y. Tanaka, S. Kashiwaya, Phys. Rev. B {\bf 56} (1997) 892.



\end{thebibliography}
\end{document}